\documentclass{resonance}
\usepackage{graphicx}
\usepackage{epsfig}
\usepackage{dcolumn}
\usepackage{bm}
\usepackage{amsmath,amssymb,color}
\usepackage{setspace}
\begin{document}

\newcommand{\nwc}{\newcommand}
\nwc{\vs}{\vspace}
\nwc{\hs}{\hspace}
\nwc{\la}{\langle}
\nwc{\ra}{\rangle}
\nwc{\lw}{\linewidth}
\nwc{\nn}{\nonumber}
\nwc{\tb}{\textbf}
\nwc{\td}{\tilde}
\nwc{\Tr}{\tb{Tr}}
\nwc{\dg}{\dagger}

\nwc{\pd}[2]{\frac{\partial #1}{\partial #2}}
\nwc{\zprl}[3]{Phys. Rev. Lett. ~{\bf #1},~#2~(#3)}
\nwc{\zpre}[3]{Phys. Rev. E ~{\bf #1},~#2~(#3)}
\nwc{\zpra}[3]{Phys. Rev. A ~{\bf #1},~#2~(#3)}
\nwc{\zjsm}[3]{J. Stat. Mech. ~{\bf #1},~#2~(#3)}
\nwc{\zepjb}[3]{Eur. Phys. J. B ~{\bf #1},~#2~(#3)}
\nwc{\zrmp}[3]{Rev. Mod. Phys. ~{\bf #1},~#2~(#3)}
\nwc{\zepl}[3]{Europhys. Lett. ~{\bf #1},~#2~(#3)}
\nwc{\zjsp}[3]{J. Stat. Phys. ~{\bf #1},~#2~(#3)}
\nwc{\zptps}[3]{Prog. Theor. Phys. Suppl. ~{\bf #1},~#2~(#3)}
\nwc{\zpt}[3]{Physics Today ~{\bf #1},~#2~(#3)}
\nwc{\zap}[3]{Adv. Phys. ~{\bf #1},~#2~(#3)}
\nwc{\zjpcm}[3]{J. Phys. Condens. Matter ~{\bf #1},~#2~(#3)}
\nwc{\zjpa}[3]{J. Phys. A: Math theor  ~{\bf #1},~#2~(#3)}

\newcommand\bea{\begin{eqnarray}}
\newcommand\eea{\end{eqnarray}}
\newcommand\beq{\begin{equation}}  
\newcommand\eeq{\end{equation}}
\newcommand{\new}{\newpage}
\newcommand{\noi}{\noindent}
\newcommand{\bib}{\bibitem}
\newcommand{\cosec}{\operatorname{cosec}}
\newcommand{\non}{\nonumber}  
\newcommand\mb{\mathbf}
\newcommand\bs{\boldsymbol}
\newcommand\mT{\mathcal{T}}
\newcommand\dd{\text{d}}
\newcommand\s{\sigma}
\newcommand{\p}{\tilde{\Psi}}
\newcommand{\ps}{\tilde{\Phi}}
\newcommand{\C}{\tilde{c}_{j}}
\newcommand{\X}{\tilde{\chi}_{{\bs k}}}
\newcommand\ie{{\it{i.e.}}}
\newcommand\etal{{\it{et al.}}}
\newcommand\eg{{\it{e.g.}}}\def\bbraket#1{\mathinner{\langle\hspace{-0.75mm}\langle{#1}\rangle\hspace{-0.75mm}\rangle}}
\def\i{\imath}
\def\v{\upsilon_F} 
\def\nn{\nonumber}
\def\f{\frac}
\def\al{\alpha}
\def\om{\omega}
\def\de{\delta}
\def\ep{\epsilon}
\def\ga{\gamma}
\def\si{\sigma}
\def\Do{\partial}
\def\De{\Delta}
\def\mb{\mathbb}
\def\mc{\mathcal}
\def\vr{\varrho}
\def\d{\cdot}
\def\t{\tilde}
\def\l{\lambda}
\def\la{\langle}
\def\ra{\rangle}
\def\mbb{\mathbb}
\def\Y{\Upsilon}
\def\ua{\uparrow}
\def\da{\downarrow}
\def\sf{\textsf}
\def\al{\alpha}
\def\be{\beta}
\def\til{\tilde}
\def\ka{\kappa}
\def\sX{\small{X}}
\def\br{{\bs r}}
\def\bk{{\bs k}}

\title{Emerging trends in Topological Insulator and Topological Superconductor}
\author{Arijit Saha and Arun M. Jayannavar}




\maketitle{}

\authorIntro{\includegraphics[width=3.5cm]{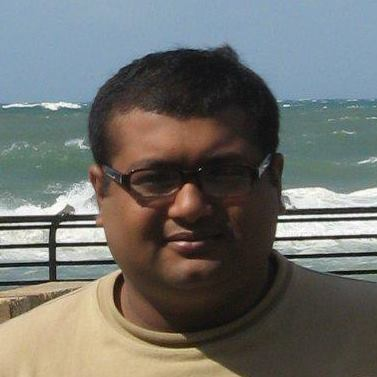}\\
Arijit Saha is a Readrer-F at Institute of Physics, Bhubaneswar. His research interest lies broadly in the 
areas of mesoscopic physics and strongly correlated electrons.
\includegraphics[width=3.5cm]{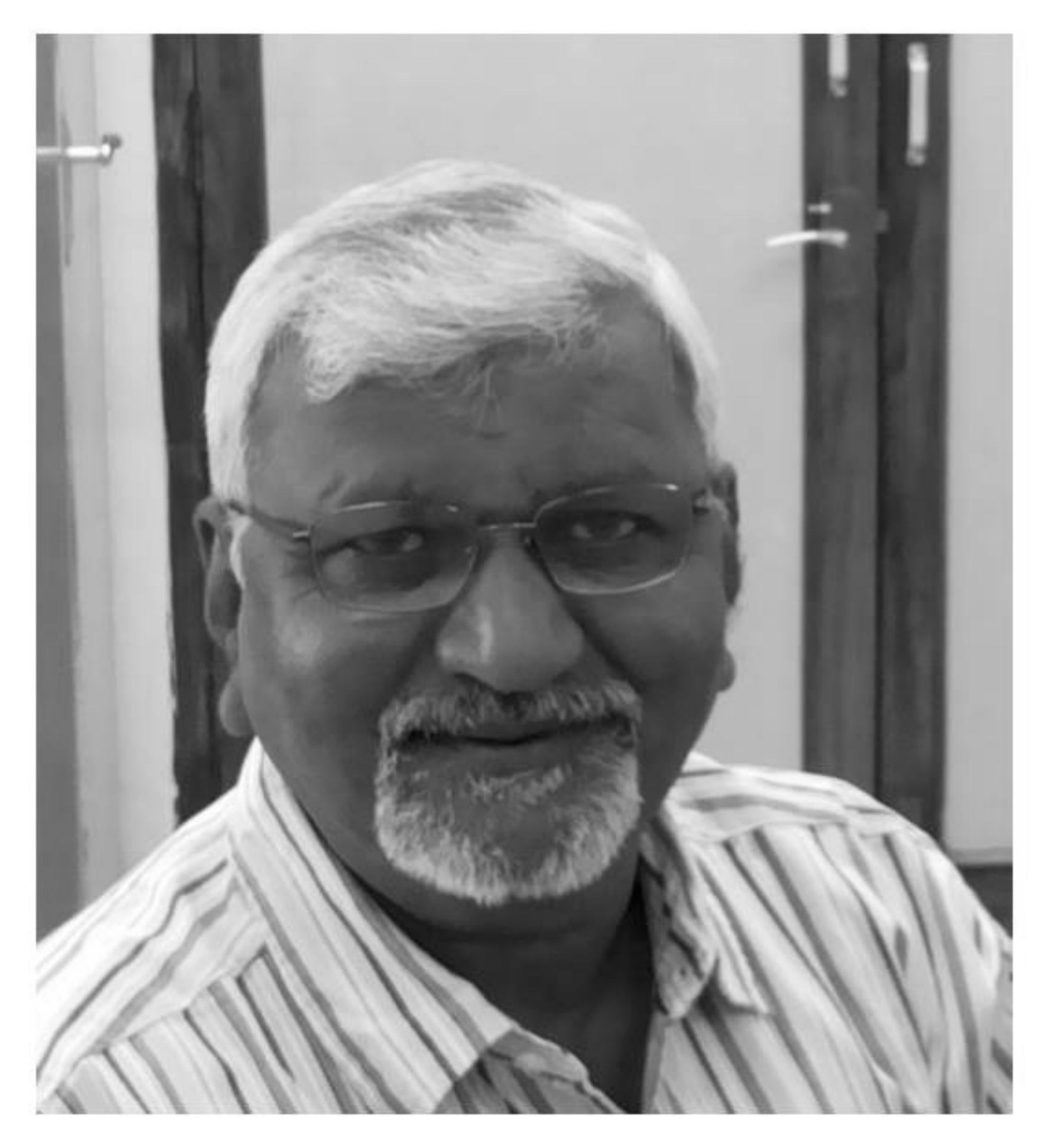}\\
Arun M. Jayannavar is a Senior Professor at Institute of Physics, Bhubaneswar. His research interest lies broadly in 
different aspects of mesoscopic physics and statistical mechanics.}

\begin{abstract}
Topological insulators are new class of materials which are characterized by a bulk band gap like ordinary band insulator but have 
protected conducting states on their edge or surface. These states emerge out due to the combination of spin-orbit coupling and 
time reversal symmetry. Also these states are insensitive to scattering by non-magnetic impurities. A two-dimensional (2D) topological
insulator has one dimensional (1D) edge states in which the spin-momentum locking of the electrons gives rise to quantum spin Hall effect. 
A three-dimensional (3D) topological insulator supports novel spin-polarized 2D Dirac fermions on its surface. These topological insulator 
materials have been theoretically predicted and experimentally observed in a variety of 2D and 3D systems, including $\rm HgTe$ quantum wells, 
$\rm BiSb$ alloys, and $\rm Bi_{2}Te_{3}$, $\rm Bi_{2}Se_{3}$ crystals. Moreover, proximity induced superconductivity in these systems can 
lead to a state that supports zero energy Majorana fermion and the phase is known as topological superconductors. In this article, the basic
idea of topological insulators and topological superconductors are presented along with their experimental developement.
\end{abstract}

\monthyear{November 2016}
\artNature{POPULAR  ARTICLE}

\section{Introduction}
In condensed matter systems, atoms with their electrons can form many different states of matter, such as crystalline solids, 
magnets and superconductors. Those states can be classified by the concept of symmetry breaking. For the above mentioned
examples, translational, rotational and guage symmetries respectively, are spontaneously broken. Before 1980, principle
of broken symmetry was the key concept for the classification of states of matter. The discovery of quantum Hall effect (QHE)
in 1980~\cite{QHExp1980} provided the first example of a quantum state where no spontaneous symmetry was broken. In QHE, the behavior 
of electrons confined in a 2D electron gas and subjected to a strong magnetic field of the order of few tesla~\cite{QHExp1980} manifests an entirely 
different topological type of order. Its behavior is independent of its specific geometry. Hence, the quantum Hall (QH) state was topologically distinct 
from all previously known states of matter. 

In recent times, a new class of topological states has emerged, called quantum spin hall (QSH) phase or topological insulator (TI)~\cite{qi2010quantum,
moore2009topological,maciejko2011,hasanmoore2010,hasan2010colloquium,sczhangreview}. Such states are topologically distinct from all other known 
states of matter, including the QH states. QSH systems are insulating in the bulk which means that they have an energy gap separating the conduction and valence 
bands. On the contrary, they have metallic edge or surface states on the boundary. These boundary edge or surface states are topologically protected 
and immune to scalar (non-magnetic) impurities. This means back-scattering is prohibited by such impurities along the edge or surface. 
Moreover, these boundary states are protected by time reversal (TR) symmetry. Here lies the important difference between a QSH state and QH state. The latter requires 
an external magnetic field which explicitly breaks the TR symmetry. QSH states, in contrast, are TR invariant and do not require an external magnetic field. 
The signatures of QSH states have been experimentally observed in $\rm HgTe$ quantum wells~\cite{konig2007quantum}, in $\rm BiSb$ alloys~\cite{dheish2008}, 
and in $\rm Bi_{2}Se_{3}$, $\rm Bi_{2}Te_{3}$ bulk crystals~\cite{yxia2009,hzhang2009}.

\section{Quantum Spin Hall Effect {\label{sec:II}}}
\begin{figure}[!thpb]
\centering
\includegraphics[width=1.0\linewidth]{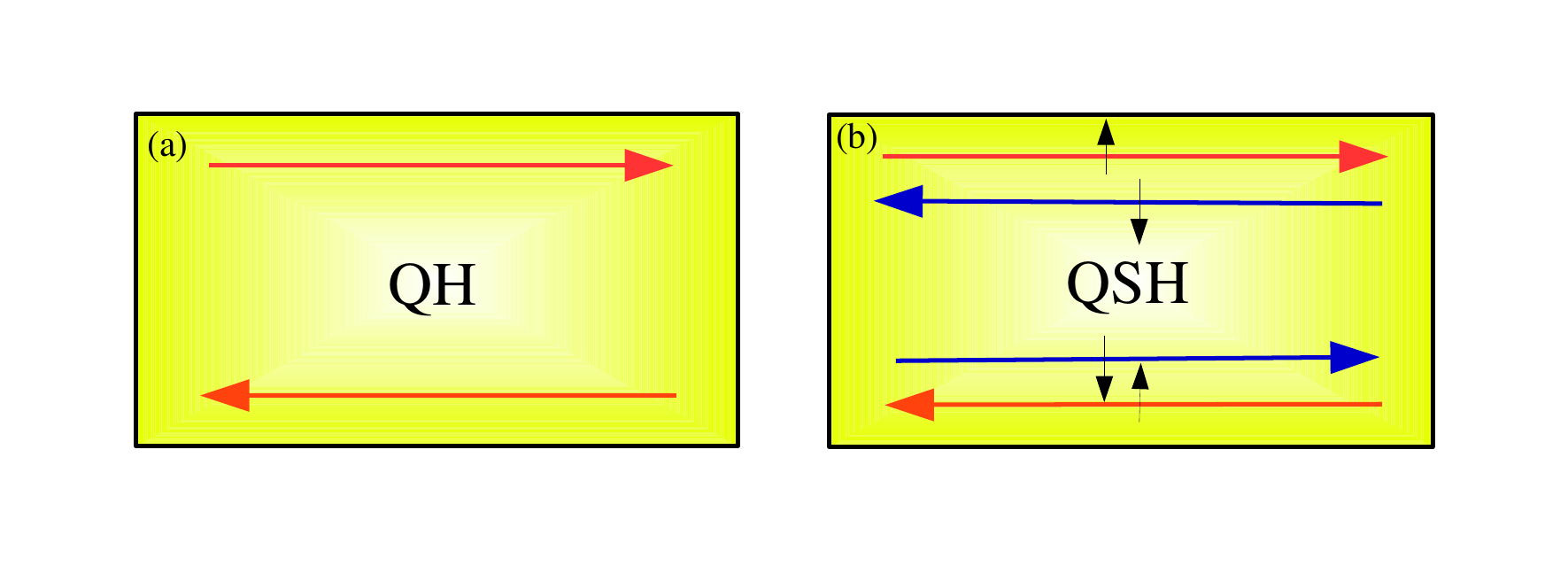}
\caption{(Color online) (a) Schematic of a QH bar in which the upper edge contains a forward mover and the lower edge contains only a
backward mover. Here both the 1D edges are spinless. (b) Cartoon of a QSH bar in which both the 1D egdes are spinful and spin-momentum locked
\ie~of helical nature. The upper edge contains a forward mover with up spin and a backward mover with down spin. The spin and momentum 
direction is reversed for the lower edge.}
\label{QSH}
\end{figure}

In QH effect, a strong magnetic field is applied perpendicular to a 2D electron gas in a semiconductor. Here magnetic field breaks the TR symmetry.
At low temperatures and high magnetic fields, the electrons flow along the edges of the 2D sample. In Fig.~\ref{QSH}(a), we present the schematic 
of a QH bar geometry where the upper and bottom 1D edges are separated by the bulk. At these two spinless 1D edges, electron propagates in a chiral 
fashion \ie~only in forward (right moving) or backward (left moving) direction. This is in contrast to normal 1D systems where electrons can 
flow in both the directions. 
Hence, the top or bottom edge of a QH bar contains only half the degrees of freedom compared to a normal 1D system. When an edge-state electron 
encounters an impurity, it still propagates along the same direction as backscattering is prohibited along the same edge. This is the key reason
why the QH effect is topologically robust. Such dissipationless transport mechanism can be very useful for semiconductor devices. 

Now one asks the following question, can we still realize a QH effect without a magnetic field \ie~without breaking TR symmetry? In recent times, 
it has been observed that certain materials with strong spin-orbit coupling (SOC) can exhibit such intriguing phenomena. SOC arises in
a material due to inversion asymmetry as well as crystal asymmetry. It is a relativistic effect and acts like an internal magnetic field
without violating the TR symmetry. 
Within such material, we can leave the spin-up forward mover and the spin-down backward mover on the upper edge. In the bottom edge, 
the spin and the associated momentum directions are reversed. This is illustrated in Fig.~\ref{QSH}(b). A system with such edge states
is said to be in a QSH state, because it has a net transport of spin forward along the top edge and backward along the bottom edge, just
like the separated transport of charge in the QH state. This phenomena is known as QSH effect which was independently predicted by
Kane-Mele~\cite{kanemele2005} and Bernevig-Huges-Zhang~\cite{bernevig2006quantum} in certain theoretical models with SOC. 

\begin{figure}[!thpb]
\centering
\includegraphics[width=0.8\linewidth]{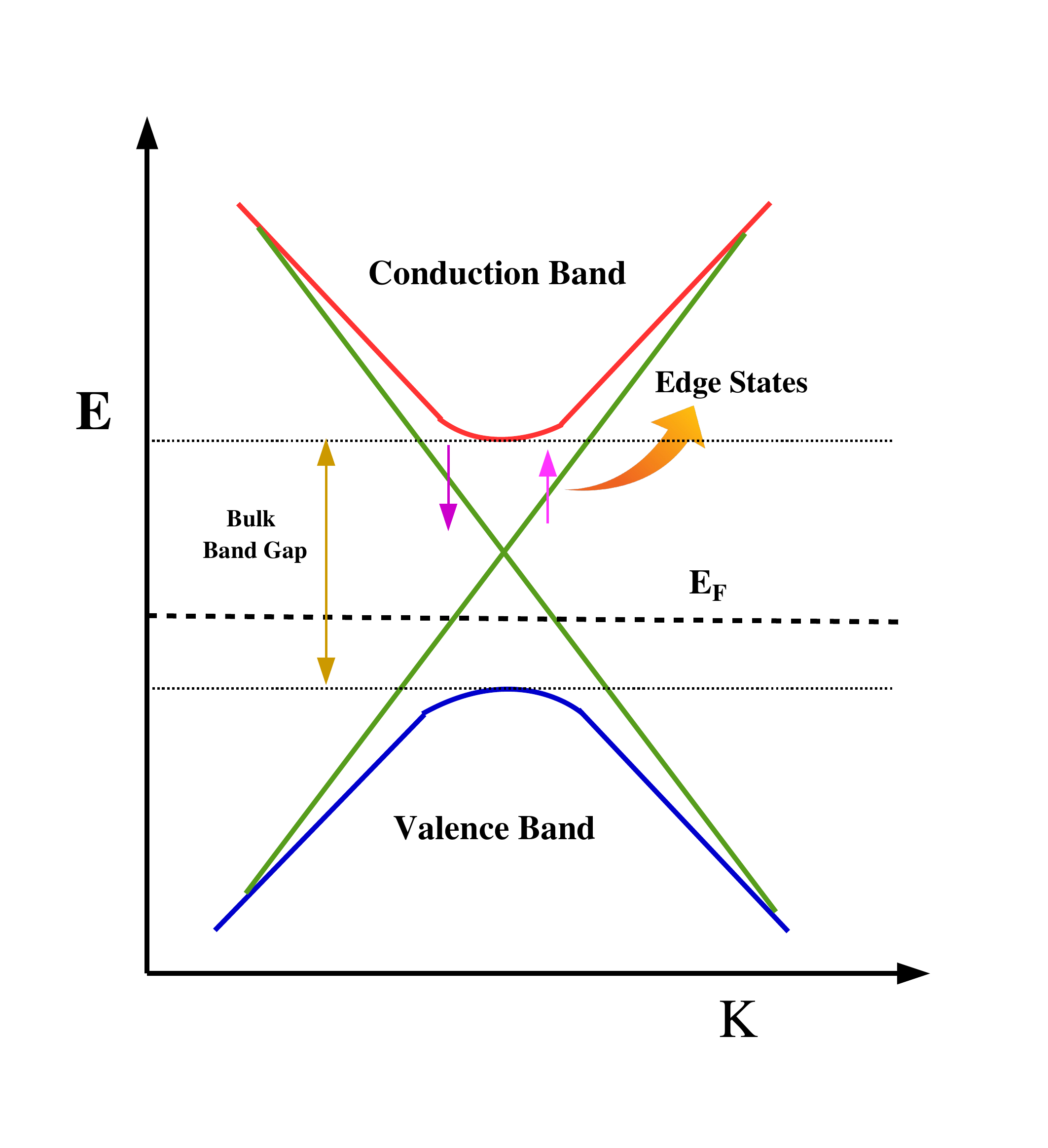}
\caption{(Color online) Schematic of the band dispersion of a 2D TI. The conduction and valence bands are separated by a bulk band gap.
The 1D helical edge states disperse linearly within the gap. Here $E_{F}$ is the Fermi energy.}
\label{band}
\end{figure}

Although QSH edges consist of both forward and backward movers, backscattering by non-magnetic impurity is still forbidden. The reason
behind this can be attributed to the fact that to have backscattering spin of the carriers also has to be flipped. Such spin-flip scattering
process is forbidden for a non-magnetic/scalar impurity. If the impurity carries a magnetic moment, then the TR symmetry is broken and
backscattering is possible due to the spin-flip process caused by the magnetic impurity. In that sense the robustness or topology of the 
QSH edge state is protected by the TR symmetry. The possibility of obtaining symmetry protected, dissipationless spin current through 
QSH systems can be very useful for future generation spintronic devices~\cite{pesinmacdonald2012}. 

\section{$\rm 2D$ Topological Insulator {\label{sec:III}}}
We already mention that SOC \ie~coupling between spin and orbital motion is a relativistic effect most pronounced in heavy elements
(elements with large Lande $\rm g$-factor). Although all materials have SOC, only few of them turns out to be topological insulators. 
Here we discuss a general mechanism for finding TI~\cite{bernevig2006quantum,maciejko2011}. It was predicted particularly for 
mercury telluride (HgTe) quantum wells which is believed to be a 2D TI. 

The typical band dispersion of a 2D TI is shown in Fig.~\ref{band}. Here the bulk conduction band and the bulk valence band is separated
by an insulating gap like a ordinary band insulator. The 1D helical edge states appear within the gap with a linear dispersion. 
The general mechanism behind the appearence of such edge states is band inversion, in which the usual ordering of conduction band and
valence band is inverted by SOC. This mechanism, we discuss next in detail, for the case of HgTe. 

\begin{figure}[!thpb]
\centering
\includegraphics[width=1.1\linewidth]{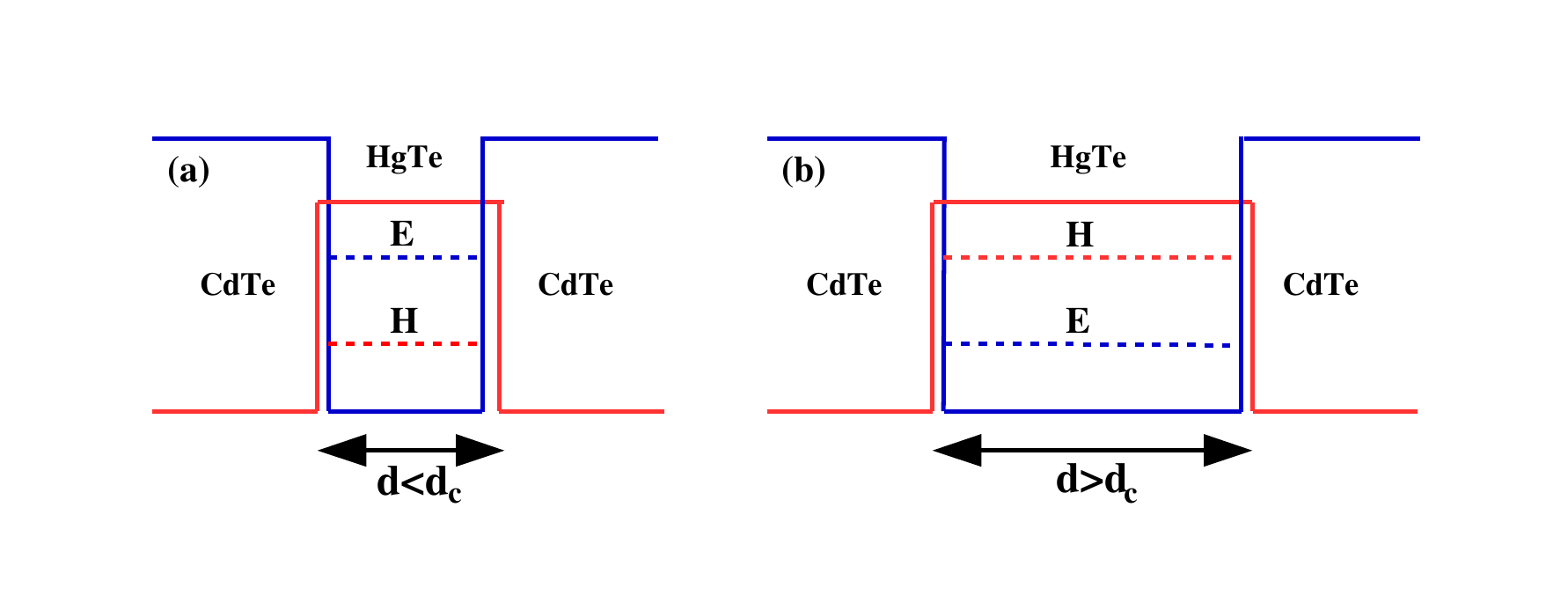}
\caption{(Color online) Cartoon of a HgTe quantum well structure sandwitched between CdTe layers. (a)~For a thin HgTe well ($d<d_{c}$),
conduction subband $\rm E$ and valence subband $\rm H$ are located like a ordinary insulator. (b)~On the other hand, in case of a 
thick HgTe well ($d>d_{c}$), the $\rm E$ and $\rm H$ subbands are inverted.}
\label{HgTe}
\end{figure}
In most common semiconductors the conduction band is formed by $s$ orbital electrons and the valence band is formed from
electrons in the $p$ orbital. However, in some heavy elements like Hg and Te, the SOC coupling is so large that the $p$ orbital 
band is pushed above the $s$ orbital \ie~the bands are inverted. HgTe quantum wells can be fabricated by sandwitching the material
between cadmium telluride (CdTe) (see Fig.~\ref{HgTe}), which owns similar lattice spacing as HgTe but weaker SOC than that. 
Therefore, as one increases the thickness $d$ of the HgTe layer, the SOC strength is enhanced for the entire quantum well.
For a thin quantum well, as shown in Fig.~\ref{HgTe}(a), the CdTe has a dominating effect and the bands follow normal ordering.
The $s$-like conduction subband $\rm E$ is located above the $p$-like valence subband $\rm H$ and the system behaves like a trivial insulator.
With the enhancement of $d$ above a critical thickness $d_{c}$, the $\rm H$ subband is pushed above the $\rm E$ subband by SOC as illustrated 
in Fig.~\ref{HgTe}(b). Due to the band inversion, a pair of gapless 1D edge states carrying opposite spins appear and they disperse linearly 
all the way from valence band to conduction band (see Fig.~\ref{band}). This pair of edge states is also known as Kramer's pair (TR partner)
and cannot be removed by external perturbations. This is one of the topological signatures of a 2D TI. 

The signature of 2D TI was observed in a recent experiment~\cite{konig2007quantum} in which they grow the HgTe quantum well by molecular
beam epitaxy method. The thickness of the HgTe layer was tuned by a gate voltage. They observed that when $d>d_{c}$ ($d_{c}\sim 6.5~\rm nm$)
\ie~the system is in the topological phase, conductance appears to be quantized ($2e^2/h$) as the two edge states of TI act as two conducting
1D channels contributing $e^2/h$ each. In contrast, when $d<d_{c}$, conductance comes out to be vanishingly small akin to trivial band insulator.

\section{$\rm 3D$ Topological Insulator {\label{sec:IV}}}
\begin{figure}[!thpb]
\centering
\includegraphics[width=0.8\linewidth]{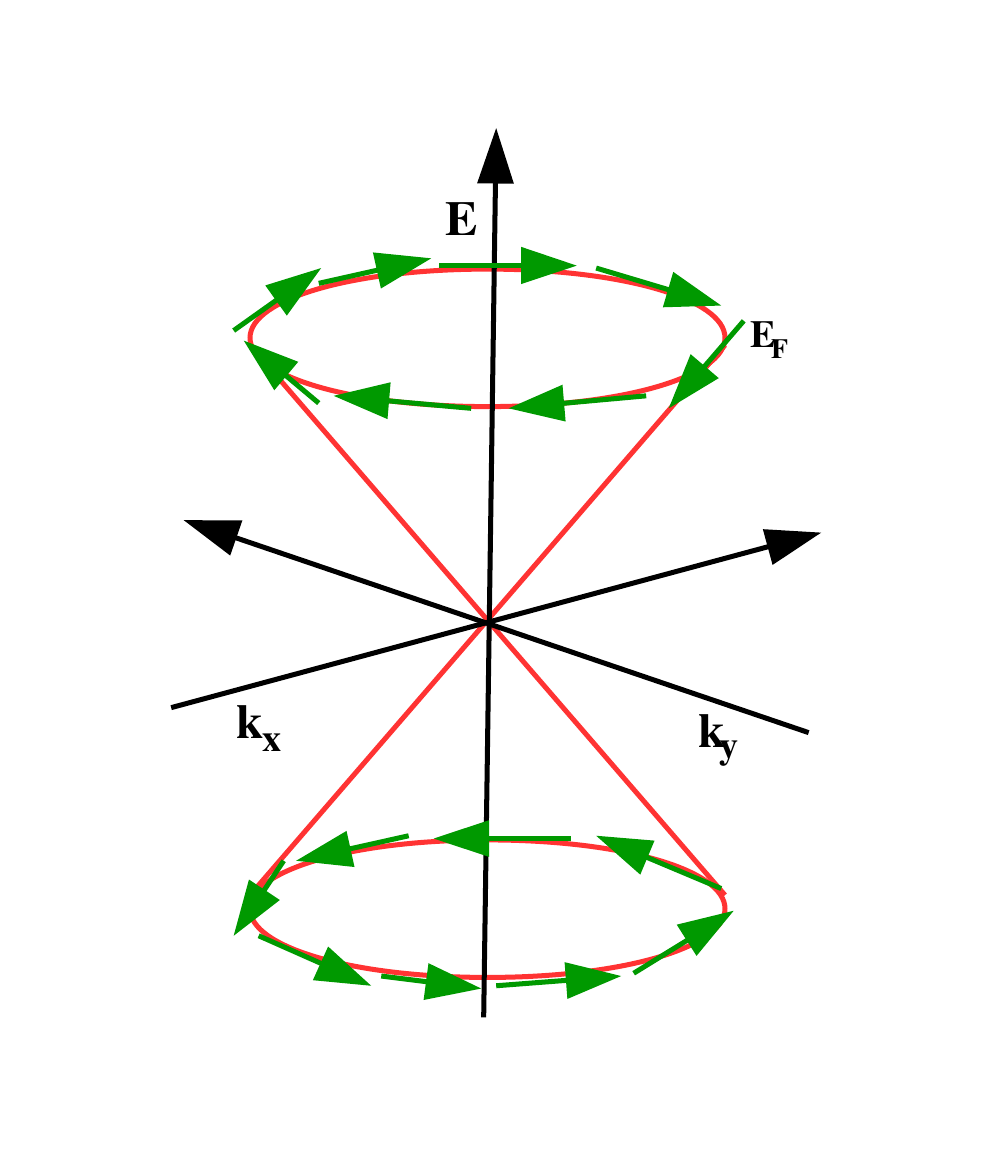}
\caption{(Color online) Schematic of the surface dispersion relation of a typical 3D TI. The 2D surface states reveal a single Dirac cone.
Rotation of the spin degree of freedom around the Fermi surface exhibits spin texture.}
\label{band3D}
\end{figure}

Here we briefly discuss the phenomenology of 3D TI~\cite{hasanmoore2010}. Note that, the pair of 1D edge states for our previous 2D TI crosses 
at $k=0$ which is already depicted in Fig.~\ref{band}. Near the crossing point, the dispersion of these states follows a linear relation. 
This is exactly the dispersion relation one obtains in relativistic quantum mechanics from the Dirac equation for a massless free fermion in 1D.
Thus the same equation can be used to describe QSH edge states. Similar picture can be generalized to a 3D TI which owns 2D surface states at
the boundary. These surface states consist of 2D massless Dirac Fermions and the corresponding dispersion forms a single Dirac cone as depicted
in Fig.~\ref{band3D}. Similar to the 2D case, the crossing point of the surface states is located at the tip of the cone \ie~$k_{x}=k_{y}=0$. 
The latter is also a TR invariant point at which Kramer's degeneracy is protected by the TR symmetry. Also note that, each momentum at the 
surface has only a single spin state at the Fermi level $E_{F}$ (spin-momentum locking), and the spin direction rotates as the momentum moves
around the Fermi surface (see Fig.~\ref{band3D}). Thus, these surface states exhibit nontrivial spin texture and carry a geometrical
Berry's phase~\cite{Berry1984} of $\pi$, which makes them topologically distinct from ordinary surface states. When disorder or scalar impurities are 
incorporated on the surface, backscattering is prohibited and the metallic surface states remain robust against disorder \ie~they don't 
become localized or gapped.  

\hskip +10cm
\fbox{\begin{minipage}{15em}
Berry's phase is a geometrical phase acquired by the wave-function of a quantum particle, subjected to a Hamiltonian depending on slowly varying (adiabatic)
time dependent parameters.
\end{minipage}}
\vskip -3cm

From materials point of view, bismuth telluride ($\rm Bi_{2}Te_{3}$), bismuth selenide ($\rm Bi_{2}Se_{3}$) are examples of 3D TI. 
These materials has been investigated experimentally using angle resolved photo emission spectroscopy (ARPES) method~\cite{yxia2009,hzhang2009}.
The single Dirac cone surface state was experimentally observed. Furthermore, spin-resolved measurements probe the spin textures of the surface
and confirms that the electron's spin indeed lies in the plane of the surface and is always perpendicular to the momentum, which is in agreement with 
the theory.

\section{Topological Superconductor {\label{sec:V}}}
Topological insulators, discussed above, are not superconductor by themselves. However, superconductivity can be induced in them via 
a process called ``proximity effect''. In this process, if a non-superconducting material is kept in close contact to a bulk superconductor,
then superconducting correlation can tunnel through the non-superconducting material upto a certain length scale depending on the 
dimension of the system and interface of the two kind of materials. 
\begin{figure}[!thpb]
\centering
\includegraphics[width=0.9\linewidth]{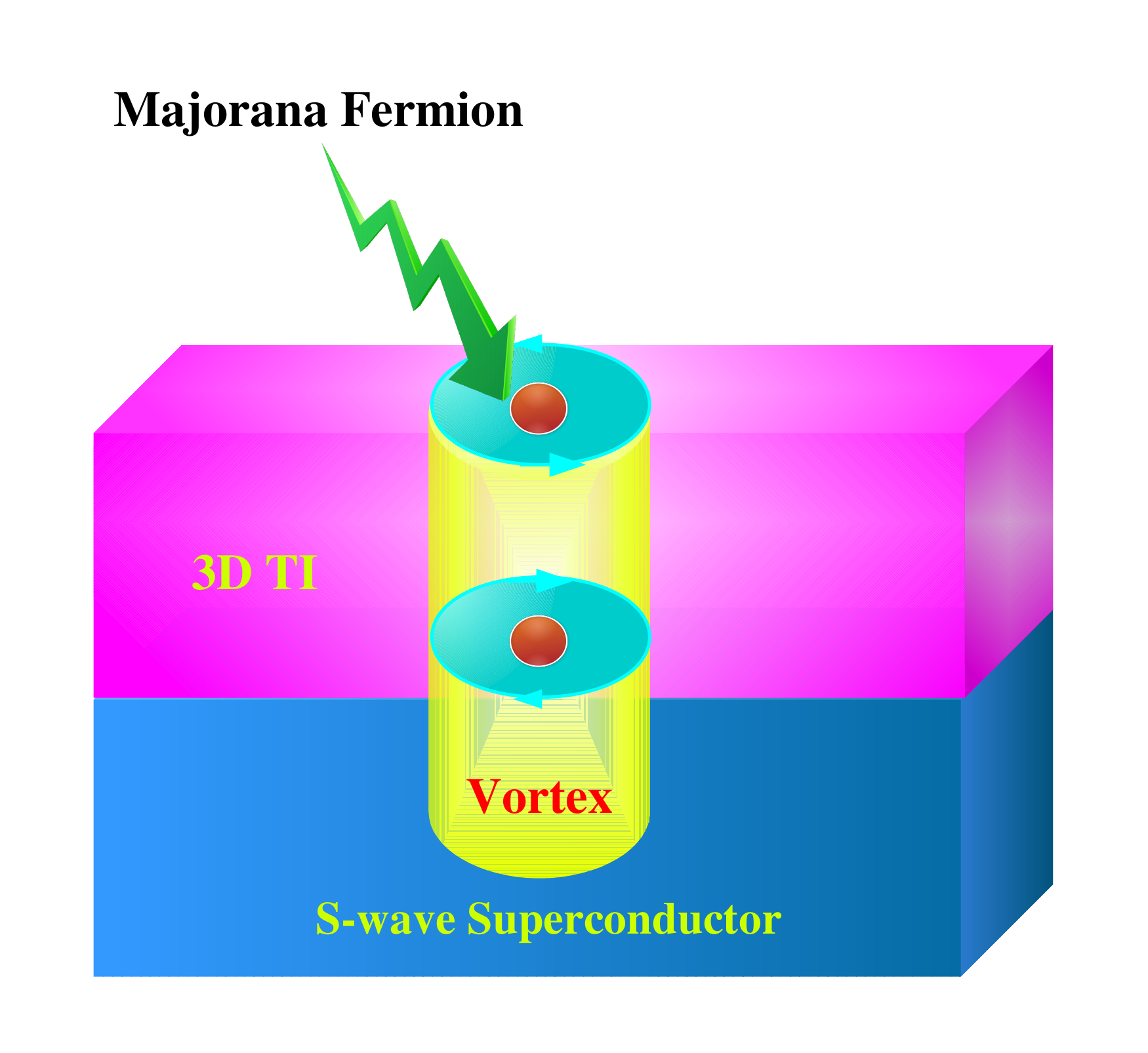}
\caption{(Color online) Schematic of a topological superconductor where a 3D TI is placed in close proximity to a conventional $s$-wave 
superconductor. Majorana Fermion is formed in the vicinity of the vortex core.}
\label{MF3D}
\end{figure}
Fu and Kane in their seminal work~\cite{fukane2008} adopted this idea and proposed that if one place a 3D TI material ($\rm Bi_{2}Te_{3}$ or $\rm Bi_{2}Se_{3}$) 
in close proximity to a ordinary superconductor,  then superconductivity can be induced in it via the proximity effect (see Fig.~\ref{MF3D}).
Moreover, the interface between a topological insulator and a superconductor may allow the creation of an ``emergent'' particle that neither 
material supports by itself. As discussed earlier, the 2D surface electrons of 3D TI are massless Dirac Fermions. Hence, a superconductor 
deposited on the surface opens an excitation gap, which can be closed locally by a magnetic field. The magnetic field penetrates as an 
Abrikosov vortex shown in Fig.~\ref{MF3D}. If a vortex line runs from the superconductor into the topological insulator, then a zero-energy 
Majorana fermion (MF) is trapped in the vicinity of the vortex core as illustrated in Fig.~\ref{MF3D}. Therefore, a proximity induced TI with
zero-energy MF is commonly known as topological superconductor (TSC). Here the MFs are revealed as the surface states of the TSC and they 
are protected by the TR and electron-hole symmetry. 

\section{Majorana Fermion {\label{sec:VI}}}
The underlying mechanism behind the emergence of MF in 3D TSC is rather a complex subject and we refer to Refs.~\cite{sczhangreview,
jalicea1,beenakkerreviewMF,stiwary1} for further details. On the contrary, we discuss here a more physical picture of MF based on 
1D systems~\cite{lejinseflensberg2012} which has been experimentally investigated very recently. To start with, one can ask the following question: 
what is a MF and what is their significance from the application point of view? MF was originally proposed by Ettore Majorana in 1937 while
finding the real solution of Dirac equation. Majorana Fermions (MFs), occurring at exactly zero energy (also known as Majorana zero modes), 
have a remarkable property of being their own antiparticles. In nanoscience and condensed-matter physics, being its own antiparticle means that 
a MF must be an equal superposition of an electron and a hole state. Mathematically, this property can be expressed as an equality between the 
particle’s creation and annihilation operators \ie~$\gamma^{\dagger}=\gamma$. Also, MFs are massless, spinless and electrically neutral. 
Furthermore, Majorana zero modes are believed to exhibit a special kind of quantum statistics, so called non-Abelian exchange statistics 
which is neither Fermi-Dirac, nor Bose-Einestein like. This special property endows MFs to be used as a building block for the future generation 
topological quantum computer, which would be exceptionally well protected from errors or decoherence~\cite{ady1}. 

\begin{figure*}
\hskip -2cm
\centering
\includegraphics[width=1.2\linewidth]{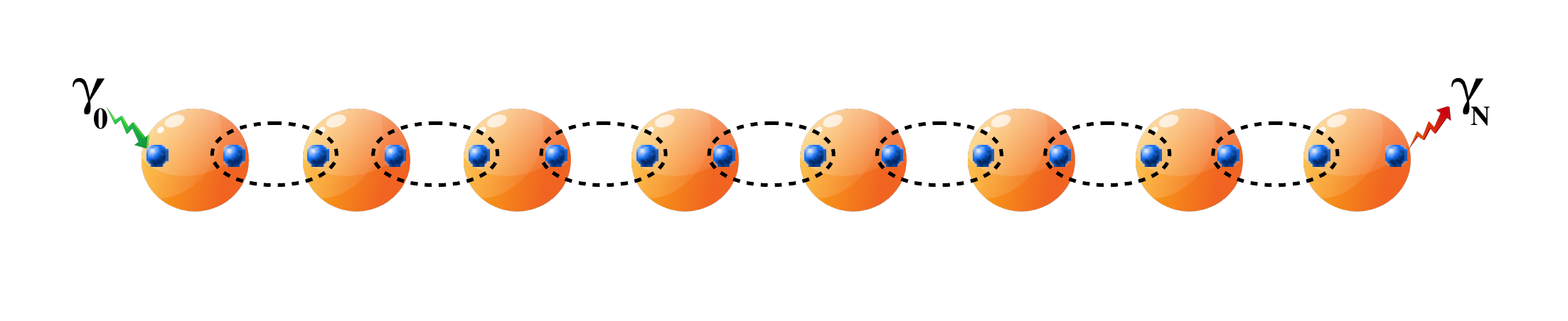}
\caption{(Color online) Cartoon of the two unpaired MFs $\gamma_{0}$ and $\gamma_{N}$, located at the two ends of the 1D chain.
The intermediate ones have paired to become Dirac Fermion.}
\label{MF1D}
\end{figure*}
The first toy model, for the realization of MF in 1D, was put forwarded by Kitaev in Ref.~\cite{kitaev}. 
In this model, one starts from a 1D tight-binding chain with unconventional spinless $p$-wave superconducting pairing.
The superconducting gap, $\Delta$, and hopping, $t$, are assumed to be the same for all lattice sites. $\mu$ is
the chemical potential set for the system. The Majorana physics can be understood for a special parameter regime 
when $t=\Delta$ and $\mu=0$. In this regime, the chain becomes TSC and two unpaired zero energy MFs $\gamma_{0}$ and $\gamma_{N}$, 
are located non-locally at the two ends of the chain. This feature is illustrated in Fig.~\ref{MF1D}. In the general case, however, 
the two MFs are not completely localized only at the two end sites of the chain, but decay exponentially away from the end. 
The MFs remain at zero energy only if the chain is long enough so that they do not overlap. For a finite chain, the two unpaired 
Majorana wave-functions can also overlap and become a normal Dirac Fermion \ie~mathematically $c=\gamma_{0}+i\gamma_{N}$
where $c$ denotes a Dirac Fermion.

\section{Realization of MF in 1D nanowire {\label{sec:VII}}}
\begin{figure*}
\centering
\includegraphics[width=1.5\linewidth]{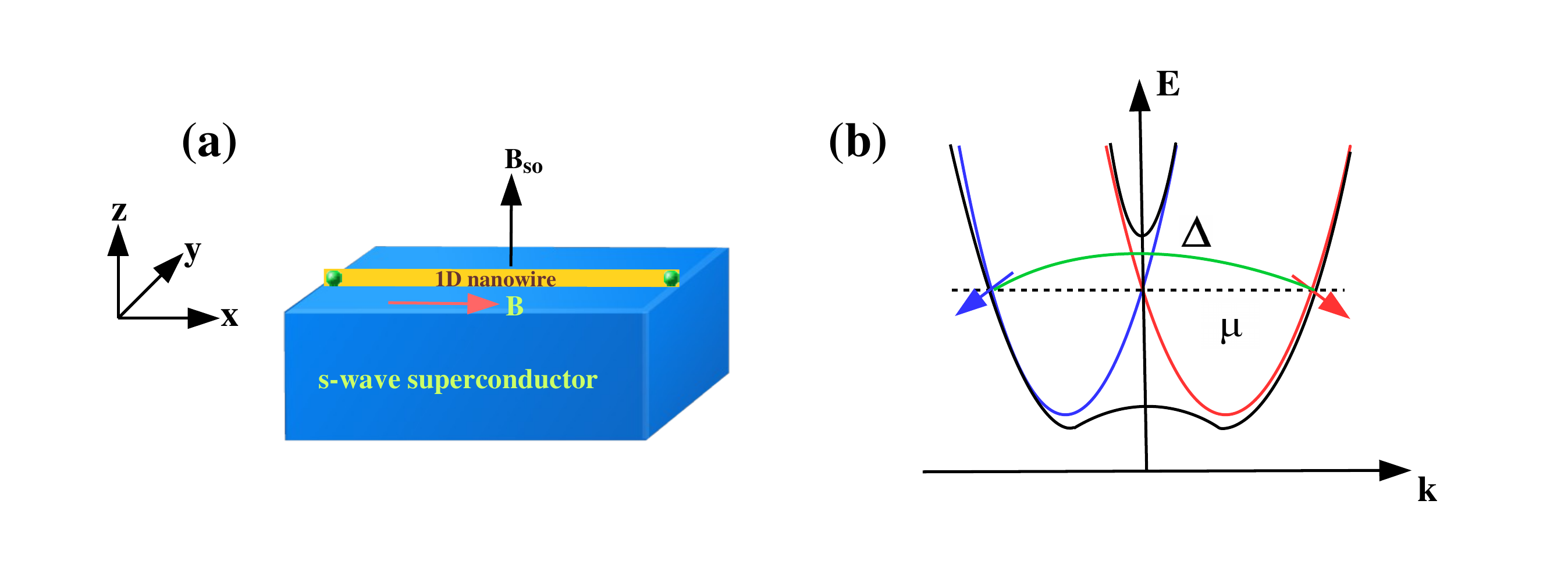}
\caption{(Color online) (a) Schematic of a 1D NW placed in close proximity to a $s$-wave superconductor. $B$ and $B_{so}$ correspond to
the homogeneous Zeeman field and internal spin-orbit field oriented perpendicular to each other. Two green dots denote the two unpaired MFs
located at the two ends of the NW. (b) Band structure of the NW in which the red and blue curves correspond to the spin-orbit splitted bands
in presence of $B_{so}$. The black curves denote the bands after the TR symmetry is broken by the Zeeman field $B$. $\Delta$ and $\mu$ are 
the proximity induced pair potential and chemical potential respectively.}
\label{MFNW}
\end{figure*}

Kitaev's chain can be realized in a 1D nanowire (NW) made of semiconductor with strong SOC. The basic idea came from two independent 
seminal works of Oreg-Refael-VonOppen~\cite{oregetal} and Lutchyn-Sau-DasSarma~\cite{jdsau2}. If a 1D NW is placed in close contact
to a conventional bulk $s$-wave superconductor, then superconductivity can be induced in the NW via the proximity effect as shown in
Fig.~\ref{MFNW}(a). A constant magnetic field $B$ is applied parallel to the NW. Here, $B_{so}$ denotes the direction of the spin-orbit
field in the NW. Note that, the direction of $B$ and $B_{so}$ should be perpendicular to each other to realize the desired phenomena. 
Under suitable circumstances, this NW becomes a TSC and a pair of MFs appears at the two ends of the NW. These two non-local Majorana
bound states are denoted by the two green dots in Fig.~\ref{MFNW}(a). 

After setting up the basic architecture required for our purpose, now we discuss how the MFs emerge in the NW.
The red and blue curves in Fig.~\ref{MFNW}(b) illustrate the band structure of the NW in the limit $B=0$. 
The strong SOC, present inside the NW, shifts the two parabolic bands depending on their spin polarization
along the axis of the spin-orbit field. Switching on a magnetic field $B$, TR symmetry is broken and a Zeeman 
gap opens up at $k=0$ which is the crossing point of the two parabolas. The new band energies are sketched 
by the black solid curves of Fig.~\ref{MFNW}(b). When the Fermi level $\mu$ resides within the gap, the wire
appears ``spinless'' (see Fig.~\ref{MFNW}(b)). 

Next, we introduce the proximity induced pairing potential $\Delta$. Hence, the gap at zero momentum \ie~$k=0$ decreases 
with the enhancement of $\Delta$ and closes completely when $B\geq\sqrt{\Delta^{2} + \mu^2}$. In this situation, the NW
enters into the topological superconducting phase. Now if we focus on the ``spinless'' regime and project away the upper
unoccupied band, then an effective intraband $p$-wave pairing mediated by $\Delta$ appears, which connects smoothly to
the phenomena demonstrated before by the Kitaev's toy model. Since the NW is in the topological superconducting phase, 
a pair of MFs emerge localized at the wire's ends (see Fig.~\ref{MFNW}(a)). For larger values of $\Delta$ the gap reopens,
but now in a non-topological superconducting state where the NW no longer appears "spinless" resulting in a trivial phase.
Therefore, the phase transition between the topological and nontopological superconducting states can only take place at
the point where the gap at $k=0$ closes. This can be achieved by satisfying a criterion $B=\sqrt{\Delta^{2} + \mu^2}$.

Very recently, the signature of MFs has been experimentally observed by Mourik \etal~\cite{VMourik} and Das \etal~\cite{Adas} 
in 1D NW system. The experiment has been performed in a set up similar to Fig.~\ref{MFNW}(a). The NW is made of 
indium antimonide ($\rm InSb$) or indium arsenide ($\rm InAs$) which has a large "$\rm g$"-factor ($\rm g\approx50$) \ie~strong SOC. 
Niobium nitride ($\rm NbN_{2}$) is used as a bulk superconductor to induce superconductivity in the NW. The signature of 
MFs is revealed via the transport measurements. When the external magnetic field, applied parallel to the NW, satisfies 
the criterion $B=\sqrt{\Delta^{2} + \mu^2}$, zero energy Majorana bound states appear at the wire's ends. The tunneling 
conductance shows a large peak (quantized to $2e^2/h$ in ideal conditions) at zero-bias when the Majorana mode is present 
and no peak when it is absent. Such zero-bias peak can be interpreted as an experimental evidence for the Majorana zero mode. 

\section{Conclusions and Outlook {\label{sec:VIII}}}
In this article, we have provided a pedagogical introduction to the exciting field of
topological insulator, topological superconductor and Majorana Fermions in condensed matter 
systems. We emphasize that a 2D TI has 1D helical edge states exhibiting QSH effect. A 3D TI
supports 2D surface sates which forms a single Dirac cone. We also discuss that these systems
can enter into a topological superconducting phase supporting Majorana Fermions at the vortex core.
Finally, we illustrate that topological superconducting phase can also be realized in a 1D nanowire
with strong SOC and proximity coupled to ordinary $s$-wave superconductor. In the topological 
superconducting phase, two non-local Majorana zero modes appear at the two ends of the NW. 
The zero bias peak, appeared in the tunneling conductance signal, reveals the experimental 
signature of MFs.   

In TI systems, one of the recent interest is to understand the effects of electron-electron
interaction in them. A TI with strong Coulomb interaction is called a "fractional topological
insulator" which is one of the current topics of research in this direction. Interface between
TI and other non-topological materials is also a subject of modern interest. In the context of
MF, the smoking gun signal of Majorana zero mode is still lacking. Further experimental investigations
is needed to explore their physical properties in detail. Moreover, there will also be a need for 
additional theoretical research works to understand the experimental findings. The final goal is 
of course to be able to control and manipulate quantum information stored in Majorana-based
qubit systems, that can be implemented for topological quantum computation.

\section{Acknowledgement}

One of us (AMJ) thanks DST, India for financial support (through J. C. Bose National Fellowship).

\end{document}